\documentclass[12pt,preprint]{aastex}
\newcommand{\beq}{\begin{equation}}
\newcommand{\eeq}{\end{equation}}

\newcommand{\hi}{H{\sc i}~}
\newcommand{\hia}{H{\sc i}}
\newcommand{\citei}[1]{\citeauthor{#1} \citeyear{#1}}

\newcommand{\kms}{km ${\rm s^{-1}}$~}
\slugcomment{}
\begin{document}

\title{Low-Velocity Halo Clouds}

\author{J.~E.~G.~Peek\altaffilmark{1},  Carl Heiles\altaffilmark{1}, M.~E.~Putman\altaffilmark{2}, Kevin Douglas\altaffilmark{3}}

\altaffiltext{1}{Department of Astronomy, University of California, Berkeley, CA 94720}
\altaffiltext{2}{Department of Astronomy, Columbia University, New York, New York 10027}
\altaffiltext{3}{Space Sciences Laboratory, University of California, Berkeley, CA 94720 now at Astrophysics Group, School of Physics, University of Exeter}

\begin{abstract}
Models that reproduce the observed high-velocity clouds (HVCs) also predict clouds at lower radial velocities that may easily be confused with Galactic disk ($|z| <$ 1 kpc) gas. We describe the first search for these low-velocity halo clouds (LVHCs) using IRAS data and the initial data from the Galactic Arecibo L-band Feed Array survey in \hi (GALFA-\hia). The technique is based upon the expectation that such clouds should, like HVCs, have very limited infrared thermal dust emission as compared to their \hi column density. We describe our `displacement-map' technique for robustly determining the dust-to-gas ratio of clouds and the associated errors that takes into account the significant scatter in the infrared flux from the Galactic disk gas. We find that there exist lower-velocity clouds that have extremely low dust-to-gas ratios, consistent with being in the Galactic halo --- candidate LVHCs. We also confirm the lack of dust in many HVCs with the notable exception of complex M, which we consider to be the first detection of dust in HVCs. We do not confirm the previously reported detection of dust in complex C. In addition, we find that most Intermediate- and Low-Velocity clouds that are part of the Galactic disk have a higher 60$\mu \rm{m}$/100 $\mu \rm{m}$ flux ratio than is typically seen in Galactic \hia, which is consistent with a previously proposed picture in which fast-moving Galactic clouds have smaller, hotter dust grains.
\end{abstract}


\section{Introduction}

High-Velocity Clouds (HVCs) are clouds of neutral Hydrogen (\hia) in or around the Galaxy that are inconsistent with Galactic rotation. Since the discovery of HVCs by \citet{Muller1963} many theories have been proposed for their origin (e.g.~\citei{oort66}), including Galactic fountain models (e.g.~\citei{Bregman1980}), material stripped from accreting satellite galaxies (e.g. \citei{Mayer06}) and formation in the cooling Galactic baryonic halo (e.g.~\citei{MB2004}). These origins can be constrained by the observed distribution of HVC metallicities, distances, morphologies, fluxes, and velocities. Indeed, many of the originally proposed origins have been disregarded as they have failed to meet some of these observational criteria. While many of these models can be made to fit the observed distribution of HVC velocities, many of them also predict some fraction of clouds indistinguishable in velocity from clouds in the Galactic disk, with $|V_{\rm LSR}| < 90$ \kms (e.g. figure 3 in \citei{wakker1991}): we call these low-velocity halo clouds (LVHCs). In particular, recent numerical simulations have shown that if HVCs are formed in a cooling Galactic halo scenario, we would expect to see an equal number of LVHCs formed in this process as HVCs (\citei{SL06}, \citei{PPS-L08}, hereafter PPS-L08). Indeed, the ratio of LVHCs to HVCs could be a powerful metric in distinguishing between these different HVC models.

\subsection{Models that Predict LVHCs}\label{mods}

As an example of the power of LVHCs to discriminate between models, we compared two very simplified models: one of clouds that inherit their velocity from the Galactic disk (the `disk-associated' model) and one of clouds that inherit their velocity from a static halo (the `halo-associated' model). These models in no way describe the physics involved in the production and evolution of LVHCs and HVCs, but rather are intended to highlight the effects of kinematically differing populations on LVHC and HVC statistics. 

The `halo-associated' model is one in which clouds randomly populate a Galacto-centric sphere of radius 50 kpc, having a normal distribution of velocities parameterized by a velocity dispersion $\sigma_H$ and an overall infall velocity towards the Galactic Center of $V_H$. The `disk-associated' model is one in which the clouds randomly populate a Galacto-centric cylinder with radius 15 kpc and $|z_{max}| = 10$ kpc, with velocities inherited from the rotation of the disk ($V = 220 {\hat \phi}~ {\rm km ~s^{-1}}$) nearest them, plus a random normal velocity distribution parameterized by $\sigma_D$, and an overall infall towards the disk of $V_D$. 

We then ran a Monte Carlo simulation in which we construct a random sample of clouds drawn from the distributions in each these models. We then `observe' the distributions from the solar position ($R_\odot = 8.5$ kpc) to determine the histogram of HVCs as a function of $V_{\rm LSR}$. We compare these histograms of HVCs to the histogram of HVCs in the updated Wakker and Van Woerden catalog \citep{WVW1991} and thereby fit each of these models by allowing only those two parameters, $V$ and $\sigma$, plus an overall scaling for each model to vary. The results of this $\chi^2$ fitting process are shown in Table \ref{toyparm}.

\begin{table}[htdp]
\begin{center}
\begin{tabular}{c c c c}
Model name & $V$ [\kms] & $\sigma$ [\kms] & $N_{\rm LVHC}/N_{\rm HVC}$ \\
\hline
\hline
Halo-associated & -80 & 79 & 0.69 \\
Disk-associated & 11 & 105 & 1.58 \\
PPS-L08 simulation & N/A & N/A & 0.76 \\
\hline
\end{tabular}
\end{center}
\caption{A table of the fitted parameters for each toy model and the corresponding ratio of LVHCs to HVCs. $V = V_H, V_D$ for Halo and Disk, respectively, similarly, $\sigma = \sigma_H, \sigma_D$. Also included is the LVHC-to-HVC ratio found in the simulations in PPS-L08.}
\label{toyparm}
\end{table}

\begin{figure*}
\begin{center}
\includegraphics[scale=.80, angle=0, trim=100 200 100 200]{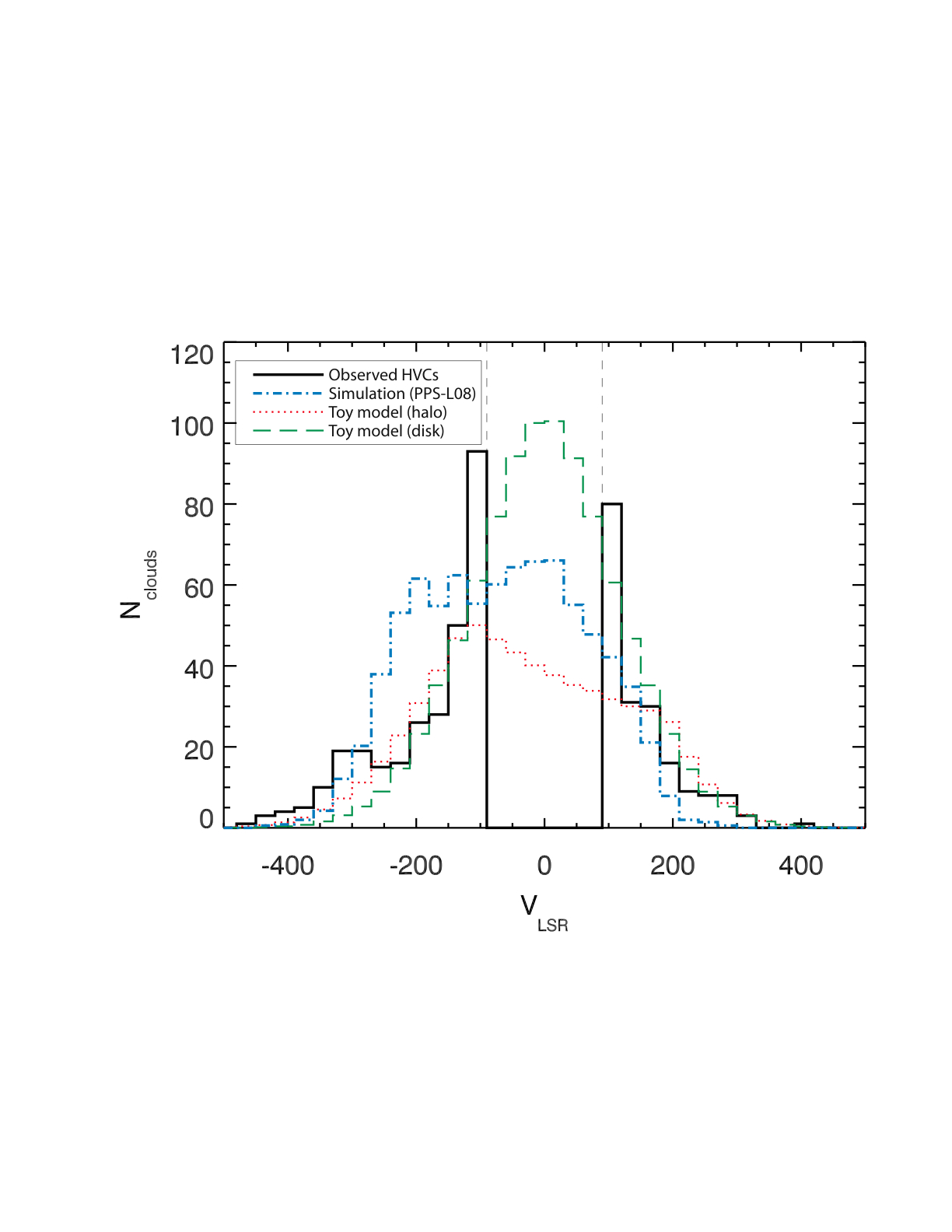}
\caption{In solid black is a histogram of HVCs by LSR velocity in the updated Wakker and Van Woerden catalog \citep{WVW1991} that fulfill the typical HVC criteria $| V_{LSR} | > 90$ \kms and $| V_{dev} | > 60$ \kms (see \citei{Wakker2004} for details). The vertical dashed lines show the region in which clouds are no longer considered HVCs by their LSR velocities. The best model fits (see \S \ref{mods} and Table \ref{toyparm} for details) are shown for the halo model (red, dotted) and for the disk model (green, long dashed). We also display the cosmological simulation data from PPS-L08 (blue, dot-dashed).
}
\label{model_plot}
\end{center}
\end{figure*}

In Figure \ref{model_plot} we show the best fit of each of these two toy models. We also show the results of the fully cosmological simulation from PPS-L08, which is not fit to the data other than an overall scaling constant to take into account the limited resolution of the simulation. We see that these two toy models and one simulation each fit the observed distribution of HVCs reasonably well, but have \emph{more than a factor of two} variation in the number of LVHCs. It is interesting to note the excess of observed clouds near the HVC cutoff velocity of 90 \kms, perhaps indicative of contamination by non-halo clouds. The variable LVHC result is consistent with plots in \citet{wakker1991} of toy models of HVCs stemming from accreting clouds and `fountain' clouds. The broad, flat distribution in the `halo-associated' model naturally stems from the `smearing out' in the velocity domain by the solar motion in the GSR frame, and therefore leads to a much lower number of LVHCs. This wide variation in LVHC number across models, ranging from 69\% to 158\% of the HVCs observed, demonstrates the possibility that LHVCs can help to distinguish among HVC origins. 

\subsection{The dust-to-gas ratio method}
One of the most important results from the \emph{Infrared Astronomical Satellite (IRAS)} was the discovery of the strong correlation between the Galactic \hi column density and the infrared diffuse emission for $|b| > 10^\circ$ (\citei{BP88}). The conclusion reached by these authors, and others, is that there is a relatively uniform distribution of interstellar radiation field (ISRF) in the Galaxy and a relatively constant dust fraction in the neutral, diffuse \hia. The ISRF heats the dust, which reradiates in the infrared, and thus infrared flux is correlated with \hi column density. As \hi is typically optically thin at high Galactic latitudes, the column density is simply proportional to the observed velocity-integrated specific intenisty. The observed ratio of infrared emission to \hi column density is called the `FIR-to-\hi ratio' or, put more simply, the `dust-to-gas ratio' (DGR), and can vary by factors of a few across the high-latitude sky \citep{BP88}. (Note that in this work we use DGR to mean the ratio of the observed intensity of dust emission to that of HI gas, rather than the ratio of their masses, as it is sometimes used in the literature.) It was also shown that HVCs \emph{do not} follow this correlation; they have no detectable dust emission (\citei{WB1986}, hereafter WB86). WB86 speculated that this paucity of emission is because either the clouds do not have very much dust in them to start with, or because they are too far away to be bathed in the ISRF of the Galaxy. In either case, this lack of infrared flux is a velocity-independent observable that is associated with HVCs and therefore also with LVHCs. We can use this effect as a discriminant between low-velocity clouds that are physically part of the Galactic disk and LVHCs. 

The aim of this paper is to use the lack of infrared flux associated with HVCs, but not other clouds, to find the first LVHCs. We also place new limits on the DGR of known HVC complexes, present a new method to determine a more accurate DGR and determine the variability in the DGR of other lower velocity clouds and one dwarf galaxy. The paper is organized as follows: in \S \ref{s_obs} we explain the observations in both infrared and \hia, in \S \ref{s_dr} the data reduction methods for each are explained, in \S \ref{s_methods} we explain our technique for determining the contribution to the infrared flux from clouds, in \S \ref{targ_sel} we describe how we choose the clouds we investigate, we give our results in  \S \ref{s_results}, and we conclude in \S \ref{s_conc}.

\section{Observations}\label{s_obs}
The observations for this project were both done in the capacity of surveys: The IRAS survey provides maps of the Galactic, diffuse mid-infrared dust emission at high latitudes and the Galactic Arecibo L-band Feed Array survey in \hi (GALFA-\hia) provides maps of the Galactic \hi in the 21-cm hyperfine transition of Hydrogen in the same region. 

\subsection{GALFA-\hi}
The radio data were obtained from the GALFA-\hi survey. ALFA is a 7-beam array of receivers mounted at the focal plane of the Arecibo 305m telescope Gregorian dome. GALFA-\hi uses this instrument to study Galactic \hi in the hyperfine transition of neutral hydrogen at 1420.405 MHz. The spectrometer, GALSPECT, has a velocity resolution of 0.18 \kms (872 Hz) and a bandwidth of 1531 km/s (7.1 MHz) centered at 0 \kms Local Standard of Rest (LSR) for each of the 7 beams and 2 polarizations. Each beam in the focal plane array has a FWHM of $\sim 3.5^\prime$, though the 6 non-central beams have significant asymmetrical sidelobes towards the outside of the array sky-footprint displaced $\sim5^{\prime}$. The GALFA-\hi survey is conducted in both simple drift and basketweave (or `meridian nodding') modes. Drift observations, taken commensally with the Arecibo Legacy Fast ALFA survey \citep{Giovanelli05}, have a typical per-beam integration time of $t_{int} \sim14.4$ seconds per beam, whereas basketweave observations have $t_{int} \sim 4.8$ seconds per beam. Some regions were observed with multiple passes in either mode, thus increasing the sensitivity of the observations. The final sky coverage of the survey will be $\sim 13000$ deg$^2$, covering the entire right ascension range from $-1^\circ$ declination to $39^\circ$ declination. At the time of writing less than $5000$ deg$^2$ of this region had been surveyed, collected between Spring 2005 and Spring 2007 in the ongoing GALFA-\hi survey. 

\subsection{IRAS}
IRAS observed the Galaxy over a period of 300 days in 1983 in the infrared wavelength bands centered on 12, 25, 60 and 100 $\mu$m \citep{Beichman87}. 98\% of the sky was surveyed, allowing relatively complete overlap with the aforementioned 21-cm line observations. Though more sensitive and higher resolution observations have since been conducted with the Spitzer Space Telescope, the publicly available IRAS data were chosen as a comparison data set for their relative completeness and comparable resolution to GALFA-\hia. In this work we only concern ourselves with the 60 and 100 $\mu$m bands, which trace the cooler dust associated with \hia.

\section{Data Reduction}\label{s_dr}
\subsection{GALFA-\hi}\label{dataHI}

To generate the \hi data cubes, raw data were reduced in the GALFA-\hi standard reduction pipeline (version 2.3), the details of which are described in \citet{PH08}. See also \citet{Stanimirovic2006} for a description of the reduction process. Corrections are made for the IF bandpass, gain variation in the the receivers, impedance mismatches in the signal chain, static RF fixed-patten noise (`baseline ripple') and overall system gain. We reduce the contamination from all of these effects significantly, but the resulting maps are not wholly without systematic effects. In \S \ref{acc} we address the effects these systematics may have on our results.

\begin{figure*}
\begin{center}
\includegraphics[scale=.70, angle=0, trim=50 25 50 100]{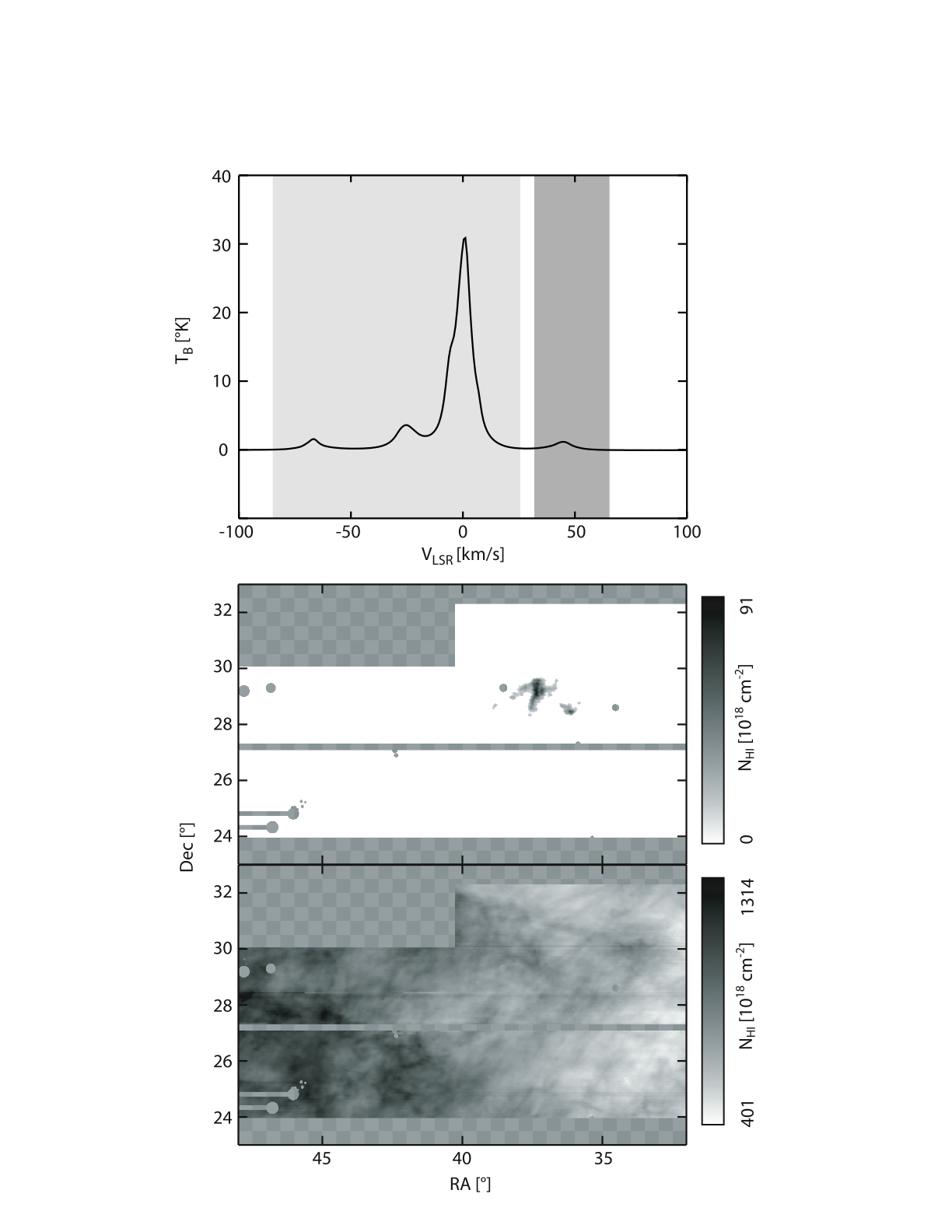}
\caption{These plots show a typical region of the sky we examine and the associated cloud and zero-velocity gas column density maps. The top plot is the average spectrum over lines of sight where we detect the L8 cloud in the \hi data. The light gray velocity range is the range over which we integrate to determine the zero-velocity gas column density; the dark gray is the range for the cloud. Note that we include all Galactic \hi along the line of sight not included in the cloud to determine the zero-velocity gas column, even if this includes some clouds that are separable from the main Galactic emission. The two maps are of the cloud column density (middle) and the zero-velocity column density (bottom). The checkerboard area represents regions that are either unobserved by GALFA-\hi or the data are omitted due to issues of data quality.}
\label{mapmake}
\end{center}
\end{figure*}

In any region of the sky under consideration, we define the `cloud' and the `zero-velocity' components of the gas. The cloud must be separated in velocity from the bulk of the \hi gas, so we can make a simple cut between the two. This means that we cannot typically inspect clouds amongst the zero-velocity gas, which limits our sample considerably. We then integrate each component over their respective velocity ranges to make 2D maps of each (see Figure \ref{mapmake}). Any data containing `glitches' are masked out. Low-amplitude striation can occur in the maps. In the case of the clouds it is typically dominated by residual baseline ripple effects not reduced by our initial data reduction. This ripple is effectively an additional spurious signal that varies slowly as a function of frequency. The striation is primarily along lines of constant declination, consistent with the drift pattern in which most of the data were taken. In this case, we reduce this effect by removing an average striation, measured in adjacent regions with the same declination. These adjacent regions contain no \hi flux, and so are good measures of the striation to be removed. The striations in the zero-velocity gas, also typically consistent with the drift pattern (constant declination), primarily stem from errors in our fits to the varying beam gains. The zero-velocity maps are cleaned by first averaging the map across declination. This declination-averaged map is then smoothed to a scale slightly larger than the striations. The smoothed component is then subtracted from the declination-averaged map to determine the striation component. This striation component is then in turn subtracted from the original zero-velocity map to yield a map with significantly fewer striations.

We also note that the Arecibo telescope is known to have non-trivial distant sidelobes (stray radiation), wherein some radiation from other parts of the Galactic \hi sky can contaminate the observations. Comparisons to the stray-radiation corrected Leiden-Argentina-Bonn survey \citep{Kalberla2005} do not show detectable differences in the shape of the HI spectra, though we do allow that some stray radiation will certainly contaminate our data and lead to increased scatter in our results.

\subsection{IRAS}
The IRAS extended emission data were originally reduced in 1984 and 1986 and were released as the SkyFlux atlas. These data were later reprocessed with more knowledge of the IRAS instrumental systematics and released as the IRAS Sky Survey Atlas (ISSA) in 1991 and 1992. This reduction, though exceedingly useful, was of limited value because it suffered from significant striping, `glitches,' zero level effects and zodiacal light. The data were once again re-reduced by \citet{SFD98} to build an all-sky extinction map. This significantly reduced the striping and `glitching' in the data. Finally, the data were re-reduced again by \citet{M-DL06} (hereafter M-DL06), increasing the effective resolution in both the 60 $\mu$m and 100 $\mu$m bands to $\sim 4^\prime$ (comparable to that of the GALFA-\hi survey) as well as fixing significant issues with zero-point offsets and zodiacal light. We use this reduction, called IRIS, because of the improved angular fidelity in the bands of interest. In addition to the reduction outlined in M-DL06, we search the data for point sources down to a limiting magnitude of 1 Jy, as we are not interested in the infrared emission from stars. We remove these point sources by fitting with a typical PSF as quoted in M-DL06. The PSF of IRAS is not simply Gaussian, so the brightest stars (typically $\leq$ 1 / field) are difficult to remove and are therefore masked out of the final comparison 60 $\mu$m and 100 $\mu$m images.

\section{Target Selection}\label{targ_sel}

We are interested in constraining the DGR for several different types of objects: HVCs, lower-velocity clouds, and Local Group dwarf galaxies. Note that the range of clouds with $|V_{\rm LSR}| < 90$ \kms has historically been divided into clouds of low velocity and clouds of intermediate velocity. As this distinction is irrelevant to this work, we will hereafter refer to all clouds distinguishable from the main Galactic zero-velocity emission with $|V_{\rm LSR}| < 90$ \kms as low-velocity clouds, or LVCs. The DGR of HVCs can be used to constrain their origin and confirm previous studies. The DGR of LVCs can be used to discern their possible membership the the Galactic halo and similarity to HVCs. The DGR of Local Group dwarfs may allow us to differentiate them from Galactic gas. Our samples of these objects are described below.

\subsection{HVCs}
Our first goal is to inspect any HVC complexes with members observed in the current GALFA-\hi data set and measure their dust-to-gas ratio. It is important to do this for two reasons. First, these new data sets and techniques yield higher fidelity assessments of the HVC DGR than have been previously made for the available HVCs, which may place tighter limits on the physical conditions of HVCs. It is also possible that we may \emph{detect} dust in these HVCs. Note that a detection of dust in HVCs, if it is at a very low level as expected, will not derail our line of reasoning for using the DGR to determine whether specific LVCs are actually HVC analogs: as long all HVCs are distinguishable from normal Galactic gas via their DGR, the reasoning holds. Secondly, it is important to establish reasonable upper limits for the DGR of HVCs to compare to our LVC sample; indeed, the main thrust of this work is to find LVCs that are similar to HVCs in their measured DGR, so a baseline upper limit for HVCs is crucial.

To find HVCs in our map we cross reference all the clouds in the WVW91 catalog with the coverage map of the current GALFA-\hi data set. We then find the highest flux cloud from each of the HVC complexes in the GALFA-\hi data set and perform image cleaning methods described in \S \ref{dataHI} on that cloud's region of the sky. In many cases we find that individual clouds in WVW91 can be resolved into multiple components, an unsurprising result given the superior resolution of the GALFA-\hi data set. For a given HVC complex, the region we use typically includes the bulk of the flux observed within the GALFA-\hi survey to date, but not necessarily the majority of the total flux of the complex. For instance, only the `tail' of complex C is visible to Arecibo, so our measurements are only of this region of complex C. We apply the `displacement map' method described in \S \ref{dmap} to the resulting maps to determined values of DGR for each cloud and associated errors.

\subsection{LVCs}
The method for finding LVCs is less straightforward than for finding HVCs, as we do not have a catalog. While some LVCs are cataloged in HVC catalogs (e.g. WVW91), it would be impossible to determine from such a lower-resolution \hi data set which are clouds separate from the zero-velocity or `disk' gas and which are intermingled with the disk gas. Given this situation, we search through the entirety of the relatively contiguous portions of the GALFA-\hi data set ($\simeq 4000^\circ$) by eye to determine likely targets for our method. These targets must have $|V_{\rm LSR}| < 90$ \kms but be distinct from the disk. We allow some very small portion of the flux to overlap from cloud to zero-velocity gas, as long it is in the far wing of the clouds, and thus does not significantly contaminate our column density maps. 

We found a total of 9 clouds for which dust measurements could be made, ranging from $|b| = 84$ to $|b| = 13$ and with fluxes ranging from 170 to 6600 Jy \kms (see Table \ref{DGR_tab}). A similar number of clouds were rejected for various reasons. These reasons include an unconstrained value of $D_{\lambda,C}$, stemming from a paucity of flux or extreme variability in the zero-velocity DGR; significant systematic errors in the GALFA-\hi data set or contamination in the IRIS data set; and mixing between the cloud and zero-velocity gas, thwarting the construction of accurate column density maps. The typical LVCs we find are relatively compact, clumpy clouds like HVCs, and are quite visible when examining the \hi data cubes. We note that this sample of clouds is not necessarily statistically representative of all LVCs.

\subsection{The Leo T Dwarf}
In addition to targeting Galactic clouds, the main thrust of this paper, we also investigated using this method to constrain the DGR of Local Group dwarf galaxies and potentially find new ones by examining the newly-discovered gas-rich dwarf galaxy Leo T. Gas-rich dwarf galaxies around the Milky Way are very difficult to distinguish in \hi from compact Galactic clouds. Leo T, for example, has a radial velocity of only 30 \kms with maximum HI intensity of 3.5K, comparable to typical Galactic local gas clouds (although relatively distinct from HVCs or LVHCs, which have a maximum HI intensity of $\simeq$ 1 K, and are typically less compact). Finding such galaxies via their HI signature alone is of great interest, as recent studies have shown that dwarf galaxies beyond 300 kpc can have very high gas fractions, and therefore may be undetectable in starlight \citep{GP08}. Observations compiled by \citet{LF98} show that dwarf irregular galaxies with HI masses below $10^{7.5} M_\odot$ have very low or undetectable DGRs. Thus, if Leo T, the only such confirmed dwarf galaxy in our observed area, does indeed have very low DGR, it is a confirmation that the same method by which we are attempting to distinguish LVHCs from other LVCs may be used to distinguish other dwarf galaxy candidates from compact Galactic HI clouds.

\section{Methods}\label{s_methods}

Since the \hi column densities of HVCs (and presumably the analogous LVHCs) are typically no more than 10\% of the high-latitude Galactic \hi column density, we expect that even in the case of a high dust-to-gas ratio cloud the infrared contribution from the cloud will be quite limited. Therefore, the methods by which we determine the dust-to-gas ratios of these clouds require significant attention to avoid contamination from systematics. 

\subsection{The simple approach}\label{stanap}
To determine the DGR of a cloud, the simplest method is to isolate the column densities from the cloud over an area of sky, $N_{C}\left(\alpha, \delta\right)$, and from the zero-velocity gas, $N_{Z}\left(\alpha, \delta\right)$, (see \S \ref{dataHI}) and solve the set of linear equations
\beq\label{simplsf}
I_{\lambda}\left(\alpha, \delta\right)  = D_{\lambda,C} N_{C}\left(\alpha, \delta\right) + D_{\lambda,Z} N_{Z}\left(\alpha, \delta\right)+ K_\lambda,
\eeq
to minimize the sum of the squares of the errors, where $I_\lambda\left(\alpha, \delta\right)$ is the infrared intensity at wavelength $\lambda$ over the region and K is an arbitrary offset in the infrared data. A least-squares solution for the equation for the range of values of $\alpha$,$\delta$ across the maps is implemented to solve for $D_{\lambda,C}$ and $D_{\lambda,Z}$, the values of the DGRs of the cloud and zero-velocity components respectively. Note that $I_{\lambda}\left(\alpha, \delta\right)$ can stem from other sources, such as dust associated with molecular gas (e.~g.~ \citei{GS06}, \citei{Gillmon06}) and dust associated with ionized gas (although see \citei{Odegard07}). These contributions are not considered in this work, and are therefore sources of error to our fits. 

We denote the `true' values of the DGRs as $D^\ast_{\lambda,C}$ and $D^\ast_{\lambda,Z}$ and the `true' offset as $K^\ast_\lambda$, in contrast to the above measured values. This technique is dependent upon the assumption that $D^\ast_{\lambda,Z}$ and $K^\ast_\lambda$ each have a single value in the region independent of position. If the zero-velocity DGR were to have significant spatial dependence within the region ($D^\ast_{\lambda,Z} = D^\ast_{\lambda,Z}\left(\alpha, \delta\right)$), $D_{\lambda,C}$ would depend critically on whether regions of high $N_{C}\left(\alpha, \delta\right)$ overlapped regions of relatively low or high $D^\ast_{\lambda,Z}\left(\alpha, \delta\right)$ as compared to the measured $D_{\lambda,Z}$. Again, it is important the realize that in this method variations in the DGR can include both variations in the true dust emissivity or fraction and variations in the contribution from other sources (molecules, ionized gas) to $I_{\lambda}$. 

\begin{figure*}
\begin{center}
\includegraphics[scale=0.7]{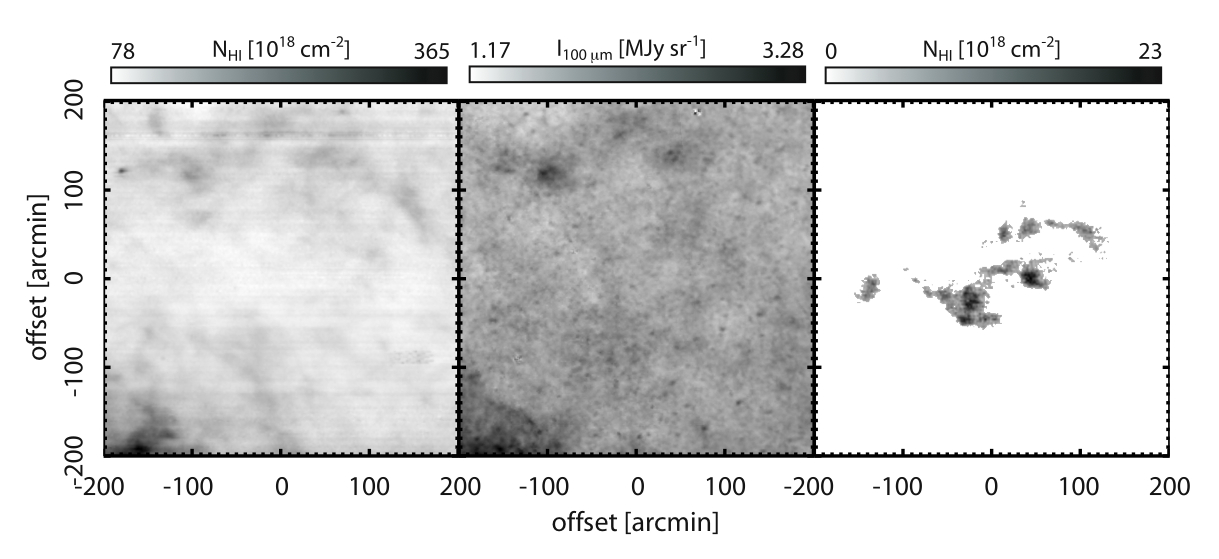}
\caption{The left image is the disk \hi column density from a region at high Galactic latitude from GALFA-\hi (l = 200, b = 50), while the center image is the 100 $\mu$m emission in the same region of sky from the IRIS data. We specifically choose a region with no gas beyond the zero-velocity disk gas, so there is no cloud-associated infrared-flux i.e. $D^\ast_{\lambda, C} = 0$. The right image is the integrated \hi column density from GALFA-\hi of a typical cloud we investigate from a different region of sky. We use these three maps as examples of $N_{Z}\left(\alpha, \delta\right)$,  $I_{100}\left(\alpha, \delta\right)$, and $N_{C}\left(\alpha, \delta\right)$, respectively to demonstrate the difficulty in using Equation \ref{simplsf} to determine $D_{\lambda, C}$.}
\label{dust_var_1}
\end{center}
\end{figure*}

\begin{figure*}
\begin{center}
\includegraphics[scale=.65]{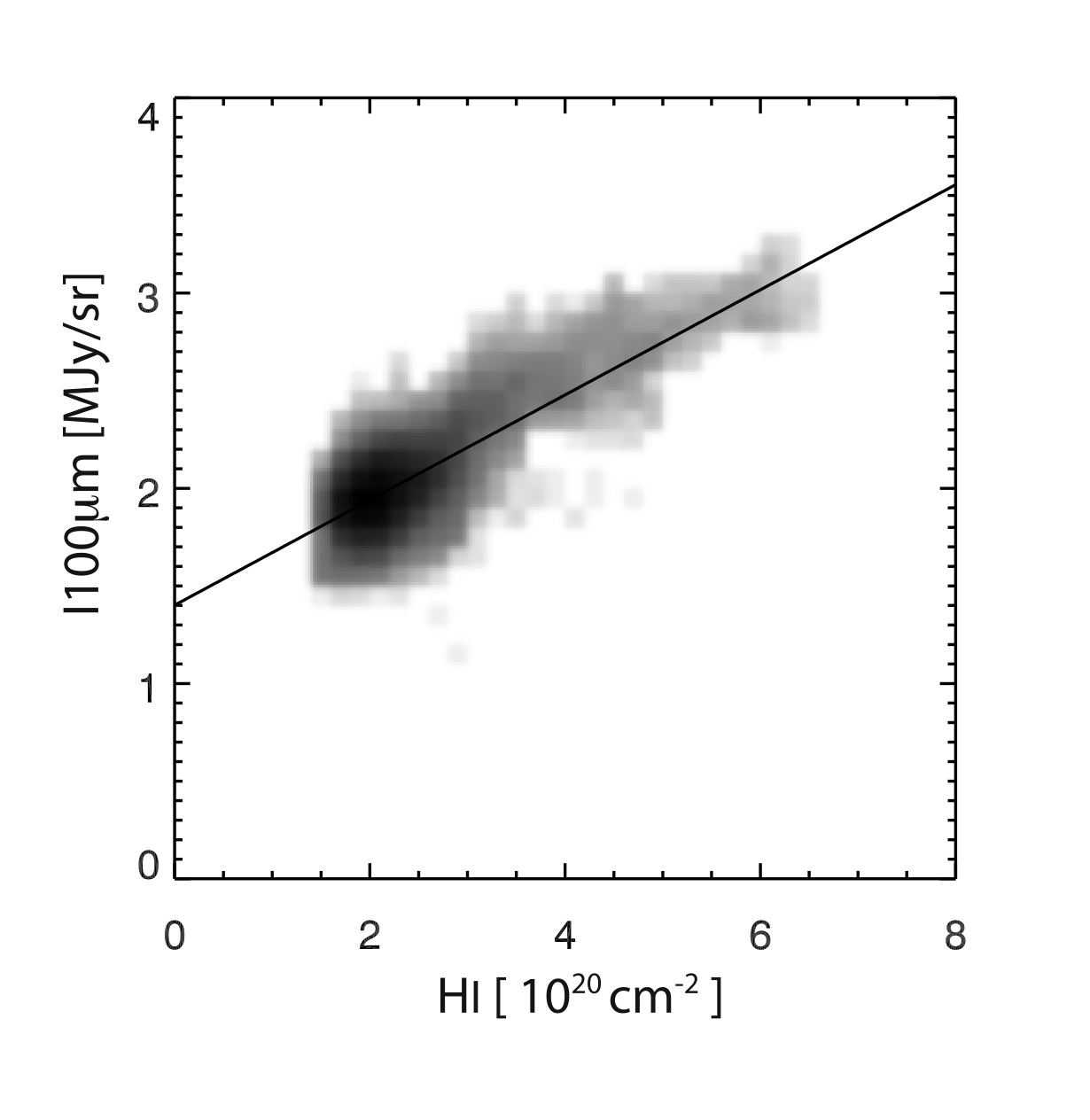}
\caption{This is a plot of the $N_{Z}\left(\alpha, \delta\right)$ vs. $I_{100}\left(\alpha, \delta\right)$ data from Figure \ref{dust_var_1}, with a superimposed linear fit to the data.}
\label{dust_var_2}
\end{center}
\end{figure*}

\begin{figure*}
\begin{center}
\includegraphics[scale=.7 ]{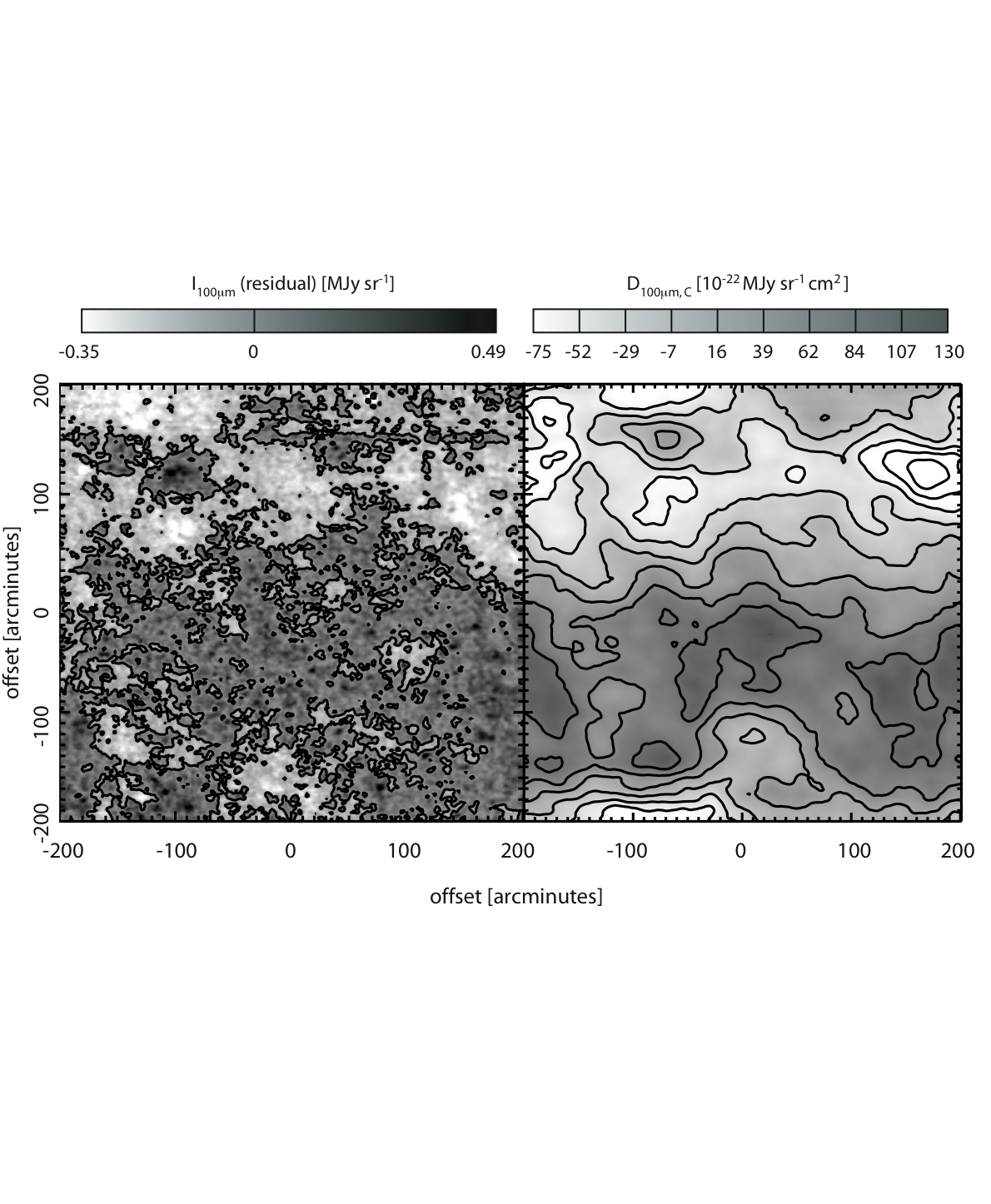}
\caption{The left image is of the residuals of the line fit shown in Figure \ref{dust_var_2}, $I_{100}\left(\alpha, \delta\right) - K_{100} - D_{100,Z} N_{Z}\left(\alpha, \delta\right)$. It is clear that there is significant spatial correlation of residuals across the region - this is typical over the regions we have examined. The right plot is the $D_{\lambda,C}\left( \Delta \alpha, \Delta \delta \right)$ `displacement map' (see \S \ref{dmap} and Equation \ref{dispmap} for details). It shows the effect of placing the cloud at different locations throughout the image $\left( \Delta \alpha, \Delta \delta \right)$ on the calculated $D_{\lambda,C}$. Therefore, the variability in this map shows the dependence of the measurement of $D_{\lambda,C}$ on the region along the line of sight to the cloud. The visual similarity between the left panel and the right panel demonstrates the effect that relatively `over-dusty' regions in the zero-velocity gas have the strong tendency to drive up the $D_{\lambda,C}$.}
\label{dust_var_3}
\end{center}
\end{figure*}

This effect is illustrated in Figures \ref{dust_var_1}, \ref{dust_var_2}, and \ref{dust_var_3}. In Figure \ref{dust_var_1} we show the integrated column density and infrared intensity from a small region of high latitude sky, column density $N_{Z}\left(\alpha, \delta\right)$ and infrared intensity $I_{100}\left(\alpha, \delta\right)$ respectively. We also show the integrated column density of a typical cloud from another region of the sky, $N_{C}\left(\alpha, \delta\right)$. Figure \ref{dust_var_2} is a plot of $N_{Z}$ vs $I_{100}$, with a line fit to the data. It is clear from this plot that there is significant scatter from a simple linear fit. This scatter has RMS noise $\sim$ 0.15 MJy sr$^{-1}$, meaning that for a cloud of typical column density $10^{19}$cm$^{-2}$ to have infrared flux equal to the scatter, it would have to have $D_{\lambda,C} \simeq 150 \times 10^{-22} \rm{MJy/sr ~cm^2}$. This residual scatter is mapped in the left panel of Figure \ref{dust_var_3}, demonstrating our claim that there is significant spatially correlated variability in $D^\ast_{\lambda,Z}$. The right panel of Figure \ref{dust_var_3} is a demonstration of the variability induced in $D_{\lambda,C}$ by moving the cloud shown in the right panel of Figure \ref{dust_var_1} around in $\alpha$ and $\delta$. There is no infrared flux associated with the cloud, but as the cloud is positioned in different locations in the sky, the  $D_{100,C}$ solution to Equation \ref{simplsf} varies dramatically. The right panel of Figure \ref{dust_var_3} is an image of this variation (see \S \ref{dmap} and Equation \ref{dispmap} for details), which is similar in form to the left panel of Figure \ref{dust_var_3}. Indeed, $D_{100,C}$ varies from $-75$ to $130 \times 10^{-22} {\rm MJy/sr ~cm^2}$, while $D_{100,Z} \simeq 35 \times 10^{-22} \rm{MJy/sr ~cm^2}$. This shows that the noise in such a measurement typically exceeds the value we are trying to measure. 

We draw the somewhat unfortunate conclusion that the application of Equation \ref{simplsf} cannot yield a believable $D_{\lambda,C}$ result in the case of the majority of Galactic clouds. This conclusion is not terribly surprising, as we know that Equation \ref{simplsf} does not include other sources of FIR radiation, and so can have significant, spatially coherent contaminants. If the cloud happens to be along the line-of-sight towards a region with particularly consistent $D_{\lambda,Z}$ and the cloud has a very high relative column density ($N_C/N_Z \sim 1$), it may be possible to use this method effectively, but for the majority of low column density clouds a new method must be developed. In the case of the claim of detection of dust in complex C using an equation equivalent to eqn. \ref{simplsf} by \citet{M-D05}, in contrast, we find complex C does not have a very high relative column density. Our analysis therefore casts significant doubt on this detection.

\subsection{The displacement map method}\label{dmap}

In the previous section we showed that displacing a cloud relative to the Galactic emission can change the measured relationship between the \hi and IR flux, the measured dust-to-gas ratio. We use this idea of displacing clouds relative to the Galactic and IR emission to develop our new measurement technique. We call this the `displacement map' method. 

	The displacement-map method is a member of a broad class of methods called `template matching', developed by the computer vision community over the last few decades (e.g.~\citei{Rosenfeld69}). In a template matching method, the goal is to determine whether, where, and to what extent an image, $f$,  is represented inside another image, $g$, or to determine if $f$ and $g$ are similar. This is done by using a `template' of $f$ and applying a wide variety of algorithms to determine the significance of a match to $g$. In computer vision the position, scale, and rotation of the template of $f$ (among other possible modifiers) are generally unknown, thus making this problem complex and computationally intensive.

	In the displacement-map method we confront a similar problem - we expect to see an IR contribution from a cloud in our infrared map, but the map contains other signals and noise as well. We expect that the IR contribution to the map from the cloud should look like the cloud (i.e.~ we assume $D_C$ is a constant); this is the template we use in the displacement-map method. As opposed to most template matching problems in the field of computer vision, we are only interested in how well the template matches at the position of the cloud on the sky. What follows is a qualitative description of the method.
	
	We quantify how well our template, the cloud, matches each position in the residual IR map after we have fitted out the dominant contribution from the zero-velocity gas. This is our displacement map, $D_{C}\left( \Delta \alpha, \Delta \delta \right)$ (note that we drop the subscript $\lambda$ in this section as the method applies to either FIR band and to reduce clutter). At the center of this map will be our `signal', contaminated by chance correlation between the residual scatter and the template. To reduce the effect of these fluctuations, we filter out large-scale structures from the displacement map -- this highlights the relatively sharp peak we expect at the center of the map. We then measure the amplitude at the center of this filtered map and the noise in that map. These measurements yield a gauge of the dust emission from the cloud and the accuracy of that result, respectively. 

Here we present the detailed, rigorous description of the method. We solve the equation
\beq\label{dispmap}
I\left(\alpha, \delta\right)  = D_{C}\left( \Delta \alpha, \Delta \delta \right) N_{C}\left(\alpha - \Delta \alpha, \delta - \Delta \delta \right) + D_{Z}\left( \Delta \alpha, \Delta \delta \right) N_{Z}\left(\alpha, \delta\right) + K\left( \Delta \alpha, \Delta \delta \right),
\eeq
for a grid of displacement values of $\Delta \alpha$ and $ \Delta \delta$, developing the displacement maps $K\left( \Delta \alpha, \Delta \delta \right)$, $ D_{C}\left( \Delta \alpha, \Delta \delta \right)$ and $D_{Z}\left( \Delta \alpha, \Delta \delta \right)$. We expect that there should be little variation across $K\left( \Delta \alpha, \Delta \delta \right)$ and $D_{Z}\left( \Delta \alpha, \Delta \delta \right)$, as the values of these fits are primarily dependent upon the correlation between $N_{Z}\left(\alpha, \delta\right)$ and $I\left(\alpha, \delta\right)$, which are independent of the variation of position of the displaced cloud on the sky, $N_{C}\left(\alpha - \Delta \alpha, \delta - \Delta \alpha \right)$. By contrast, $D_{C}\left( \Delta \alpha, \Delta \delta \right)$ should have strong variations depending upon the strength of any cloud infrared emission signal in $I\left(\alpha, \delta\right)$ and the uncompensated spatial variations in the values of $K^\ast$ and $D^\ast_{Z}$ (as shown in Figure \ref{dust_var_3}). 

Our new displacement map technique depends upon the fact that if a cloud does indeed have associated dust flux it will have a predictable effect upon the $D_{C}\left( \Delta \alpha, \Delta \delta \right)$ displacement map. We take the infrared emission to be decomposed into two parts:
\beq
I\left(\alpha, \delta\right) \equiv I^Z\left(\alpha, \delta\right) + I^C\left(\alpha, \delta\right)
\eeq
where 
\beq\label{icdef}
I^C\left(\alpha, \delta\right) = D^\ast_{C} N_{C}\left(\alpha, \delta\right)
\eeq
and $I^Z\left(\alpha, \delta\right)$ is the component of the flux associated with the low velocity gas and residuals. Note that we use the superscript C or Z to denote that the quantity is associated with infrared flux from either the cloud or zero-velocity component, whereas the subscript C or Z is used to denote whether the quantity is associated with the \hi column from either the cloud or zero-velocity component. As an example, $D_{C}\left( \Delta \alpha, \Delta \delta \right)$ represents the entire displacement map of interest, while $D^C_{C}\left( \Delta \alpha, \Delta \delta \right)$ is the component of the displacement map associated with the flux that comes from the cloud itself, $I^C$. This construction allows us to determine this influence of cloud-associated dust emission on $D_{C}\left( \Delta \alpha, \Delta \delta \right)$.

Since we solve Equation \ref{dispmap} with a linear least-squares fit, the solution can be decomposed into two solutions, one for each of the components of $I\left(\alpha, \delta\right)$. We are interested in the solution to the $I^C$ component of the equation:
\begin{eqnarray}\label{withcloud}
I^C\left(\alpha, \delta\right) - K^C\left( \Delta \alpha, \Delta \delta \right) &=& D^C_{C}\left( \Delta \alpha, \Delta \delta \right) N_{C}\left(\alpha - \Delta \alpha, \delta - \Delta \delta \right) \\
&& + D^C_{Z}\left( \Delta \alpha, \Delta \delta \right) N_{Z}\left(\alpha, \delta\right). \nonumber
\end{eqnarray}
Since to first approximation $N_{Z}\left(\alpha, \delta\right)$ should be uncorrelated to $I^C\left(\alpha, \delta\right)$ in general ($D^C_{Z}= 0$) and $I^C\left(\alpha, \delta\right)$ has no `offset' component, Equation \ref{withcloud}
can be reduced to
\beq\label{twiddle}
I^C\left(\alpha, \delta\right) \simeq D^C_{C}\left( \Delta \alpha, \Delta \delta \right) N_{C}\left(\alpha - \Delta \alpha, \delta - \Delta \delta \right)
\eeq
Substituting in Equation \ref{icdef} we find 
\beq
D^\ast_{C} N_{C}\left(\alpha, \delta\right) = D^C_{C}\left( \Delta \alpha, \Delta \delta \right) N_{C}\left(\alpha - \Delta \alpha, \delta - \Delta \delta \right).
\eeq
This is equivalent to a simple least-squares fit of slope with y-intercept = 0, whose solution is simply
\beq\label{finsol}
D^C_{C}\left( \Delta \alpha, \Delta \delta \right) = D^\ast_{C}\frac{\sum N_{C}\left(\alpha, \delta\right) N_{C}\left(\alpha - \Delta \alpha, \delta - \Delta \delta \right)}{\sum N_{C}\left(\alpha, \delta\right) N_{C}\left(\alpha, \delta\right)}.
\eeq
The ratio of the sums is simply the normalized two-dimensional auto-correlation function of $N_{C}\left(\alpha, \delta\right)$:
\beq
R_{n}\left(N_{C}\left(\alpha, \delta\right)\right) \equiv \frac{\sum N_{C}\left(\alpha, \delta\right) N_{C}\left(\alpha - \Delta \alpha, \delta - \Delta \delta \right)}{\sum N_{C}\left(\alpha, \delta\right) N_{C}\left(\alpha, \delta\right)}
\eeq
In practice, $R_{n}\left(N_{C}\left(\alpha, \delta\right)\right)$ varies much more rapidly near the origin than the overall variability of $D_{C}\left( \Delta \alpha, \Delta \delta \right)$, which typically varies relatively smoothly (e.g. the right plot in Figure \ref{dust_var_3}). We can remove this smooth variability from $D_{C}\left( \Delta \alpha, \Delta \delta \right)$ by subtracting a smoothed (i.e. convolved) version of the displacement map:
\beq
U\left(D_{C}\left( \Delta \alpha, \Delta \delta \right)\right) \equiv D_{C}\left( \Delta \alpha, \Delta \delta \right) - \kappa\left( \Delta \alpha, \Delta \delta \right) \ast D_{C}\left( \Delta \alpha, \Delta \delta \right),
\eeq
where $\kappa\left( \Delta \alpha, \Delta \delta \right)$ is our Gaussian smoothing kernel. This is equivalent to either applying an `unsharp mask' or filtering out low spatial frequency modes in the Fourier domain. While this unsharp masking will have the effect of diminishing the contribution from spurious large-scale fluctuations in the displacement map, it also somewhat diminishes the strength of the signal of interest, $D^C_{C}\left( \Delta \alpha, \Delta \delta \right)$, at the origin as well. Since unsharp masking is a linear operation,
\beq
U\left(D_{C}\left( \Delta \alpha, \Delta \delta \right) \right) = U\left(D^Z_{C}\left( \Delta \alpha, \Delta \delta \right) \right) + U\left(D^C_{C}\left( \Delta \alpha, \Delta \delta \right) \right).
\eeq
Using Equation \ref{finsol} we then solve for $D^\ast_{C}$:
\beq
D^\ast_{C} = \frac{U\left(D_{C}\left( \Delta \alpha, \Delta \delta \right) \right)\big|_{\Delta \alpha = 0, \Delta \delta = 0} - U\left(D^Z_{C}\left( \Delta \alpha, \Delta \delta \right) \right)\big|_{\Delta \alpha = 0, \Delta \delta = 0}} {U\left(R_{n}\left(N_{C}\left(\alpha, \delta\right)\right)\right)\Big|_{\Delta \alpha = 0, \Delta \delta = 0}}.
\eeq
All of the right side of this equation is known except $U\left(D^Z_{C}\left( \Delta \alpha, \Delta \delta \right) \right)$, which we ignore as small and take to be contribution a to the error.

To determine the error in this measurement, we simply calculate the standard deviation (RMS noise) in this scaled, unsharp-masked displacement map away from the origin:
\beq
\sigma = \sqrt{\Bigg\langle\left( \frac{U\left(D_{C}\left( \Delta \alpha, \Delta \delta \right) \right)} {U\left(R_{n}\left(N_{C}\left(\alpha, \delta\right)\right)\right)\Big|_{\Delta \alpha = 0, \Delta \delta = 0}}\right)^{2}\Bigg\rangle}
\eeq

We set the size of the Gaussian smoothing kernel, $\kappa\left( \Delta \alpha, \Delta \delta \right)$, to a FWHM = $10^\prime$. This size is chosen simply to maximize the signal-to-noise. This displacement map method allows us to take into account accurately the effect of the errors in our fits to the zero-velocity gas without attempting to fully parameterize them. 

\begin{figure*}
\begin{center}
\includegraphics[scale=.7, angle=0]{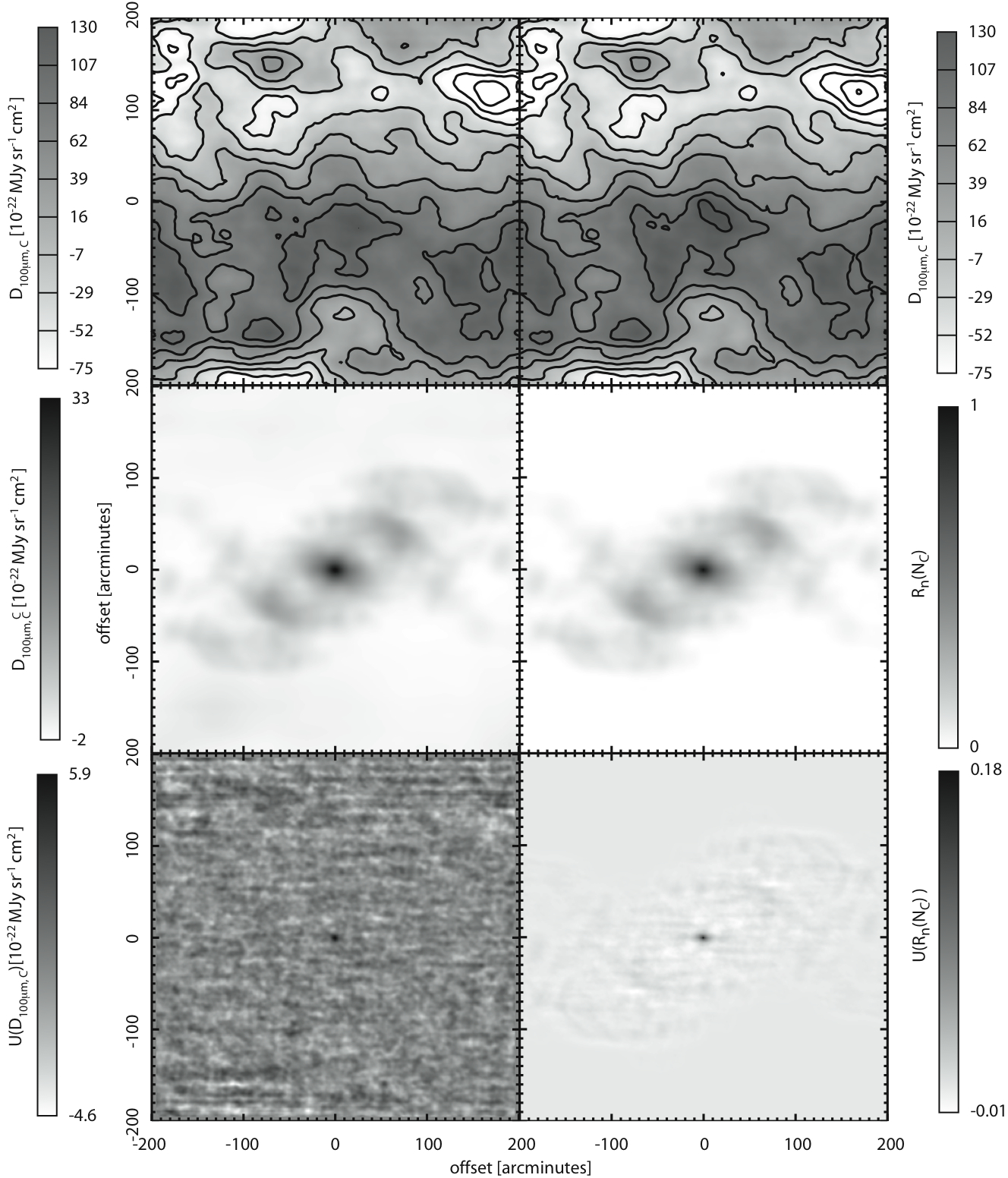}
\caption{Six plots on the same scale of one arcminute per pixel, demonstrating the displacement map method. See the end of \S \ref{dmap} for details.}
\label{usm}
\end{center}
\end{figure*}

The displacement map method is illustrated in Figure \ref{usm}. The six plots in Figure \ref{usm} are all on the same scale of one arcminute per pixel. The upper left plot is a reproduction of the right plot from Figure \ref{dust_var_3}, $D_{C}\left( \Delta \alpha, \Delta \delta \right)$ in Equation \ref{dispmap}, shown here for direct comparison to the next plot. The plot in the upper right is similar to the previous plot but of $D_{C}\left( \Delta \alpha, \Delta \delta \right)$ as in Equation \ref{withcloud}, with $D^\ast_{C} \ne 0$ by construction.  In this case we set $D^\ast_{C} = 36 \times {\rm 10^{-22}~MJy~sr^{-1} ~cm^2}$. The maps are very similar, except a very slight increase in intensity at the center of the map, corresponding to the detection of the cloud dust. The middle left plot is the difference between the two upper plots, which is to say $D_{C}\left( \Delta \alpha, \Delta \delta \right) - D^Z_{C}\left( \Delta \alpha, \Delta \delta \right) = D^C_{C}\left( \Delta \alpha, \Delta \delta \right)$. The middle right plot is of $R_{n}\left(N_{C}\left(\alpha, \delta\right)\right)$. It is empirically evident that the assumptions required to state Equation \ref{twiddle} are justified in this case, as $D^C_{C}\left( \Delta \alpha, \Delta \delta \right)$ is extremely similar to $R_{n}\left(N_{C}\left(\alpha, \delta\right)\right)$. The bottom left plot is of $U\left(D_{C}\left( \Delta \alpha, \Delta \delta \right)\right)$. This shows how the central spike that comes from $D^C_{C}\left( \Delta \alpha, \Delta \delta \right)$ can be measured in the noisy background. The right plot is of $U\left(R_{n}\left(N_{C}\left(\alpha, \delta\right)\right)\right)$, which we compare to the bottom left plot to determine $D_{C}$.

\subsection{Tests of the displacement map method}\label{tests}

Here we test the efficacy of the displacement-map method and the scale of a few possible systematic contaminants.

\subsubsection{Measurement accuracy}\label{acc}
To test the claim in \S \ref{dmap} that we can accurately measure the DGR in clouds (as well as the error in that value), we construct a numerical experiment. As in Figures \ref{dust_var_1} and \ref{usm}, we find an arbitrary high-latitude region of the sky containing no clouds and measure the $N_{Z}\left(\alpha, \delta\right)$, as well as $I^Z_{\lambda}\left(\alpha, \delta\right)$. We then find a cloud from a different region of sky and use it as our $N_{C}\left(\alpha, \delta\right)$ and assume a $D^\ast_{\lambda,C}$. We construct the dust map, $I_{\lambda}\left(\alpha, \delta\right) = I^Z_{\lambda}\left(\alpha, \delta\right) +D^\ast_{\lambda,C}N_{C}\left(\alpha, \delta\right)$. We then follow the displacement map method to measure $D_{\lambda,C}$ and the associated errors. We repeated this experiment with three different regions of the sky, three different measured clouds, each in 16 different positions, using 4 different values for $D^\ast_{100,C}$: 15, 45, 75, and 150 $\times {\rm 10^{-22}~MJy~sr^{-1} ~cm^2}$. We find using these 384 experiments that the values predicted by the method are consistent with the input value of $D^\ast_{\lambda,C}$ to better than 15\% of the predicted 1-$\sigma$ errors and that the predicted 1-$\sigma$ errors themselves are accurate to better than 5\%.

\subsubsection{Gain and baseline ripple effects}
In section \ref{dataHI} we described our methods for reducing the effects of baseline ripple and gain fluctuations. While these methods are successful in dramatically reducing these instrumental effects, the residuals of both of these contaminants can produce visible effects in the zero-velocity and cloud maps. In both the case of residual gain fluctuations and residual baseline ripple, small correlations are induced between the measured column density of the zero-velocity map and the cloud map. This correlation can have the consequence of reducing the measured $D_{\lambda,C}$. 

We test these effects by conducting numerical experiments similar to those conducted in \S \ref{acc}. We find a baseline result by using the displacement map method to determine $D_{\lambda,C}$ for both the $I_{\lambda}\left(\alpha, \delta\right)$ and the $I^Z_{\lambda}\left(\alpha, \delta\right)$ dust map. The results are consistent with the inputs, i.e. the derived $D_{\lambda,C}$ for the $I^Z_{\lambda}\left(\alpha, \delta\right)$ map is consistent with zero and the derived $D_{\lambda,C}$ for the $I_{\lambda}\left(\alpha, \delta\right)$ is consistent with $D^\ast_{\lambda,C}$. While the $N_{Z}\left(\alpha, \delta\right)$ and  $N_{C}\left(\alpha, \delta\right)$ are already contaminated by these effects, they are uncorrelated, and so should only add to the noise, rather than biasing the results of the fit. To add in this correlated noise, we then examine another region of sky for typical baseline ripple effects and typical gain effects. We apply each of these effects independently to the \hi column density maps, and re-run the displacement map technique for each set of contaminations. In each case we see an effect on the final $D_{\lambda,C}$ of $\simeq 5 \%$, in the expected sense of decreasing the effect of the fit. Since it is difficult to determine the exact amplitude of this effect in the case of each cloud we consider, we add a $10 \%$ systematic error to account for this effect.

\subsubsection{Fit region effects}
In previous sections we have referred to the `region' over which the equations are evaluated, spanning some range in $\alpha$ and $\delta$. We now ask how sensitive our solution for $D_{\lambda,C}$ is to variation in the size of that region. An experiment similar to the above section was conducted, wherein instead of adding noise to the maps, we vary the area of the region under consideration, keeping the size of the cloud fixed. We find that varying the area of the regions by factors of 3 has no detectable effect on the accuracy of the method. 

\subsubsection{Effects of fitting routines}
The standard least-squares or chi-square fit only takes into account errors in one variable in the problem. To take fully into account errors in all dimensions, one can employ Jefferys' method, an iterative chi-square technique \citep{Jefferys80}. Unfortunately, this method can be somewhat unstable and is strongly sensitive to the assumed errors and covariances, which may be difficult to determine robustly. The method can also be very computationally intensive, particularly when being employed many times as described in \S \ref{dmap}. For these reasons we wish to determine the sensitivity of our results to the methods by which we solve our linear equations, and specifically whether we get accurate results using a standard least-square fit. We performed tests similar to those in the previous sections, but varied our fitting method, alternately using $I_{\lambda}\left(\alpha, \delta\right)$ or $N_{Z}\left(\alpha, \delta\right)$ as our independent variable, assumed to have no errors. While the values measured by the `simple approach'  (\S \ref{stanap}) did vary significantly between fixing different independent variables, and did not match our input $D^\ast_{\lambda,C}$, our results using the displacement map method were effectively identical to each other and matched $D^\ast_{\lambda,C}$ to within the measured noise. Having empirically shown that the more complex multi-error methods are unnecessary, we use the standard least-squares fit method to determine $D_{\lambda,C}$.

\begin{center}
\begin{deluxetable}{lrr rr rrr}
\tablewidth{0pt}
\tablehead{
\colhead{Name} & \colhead{l} & \colhead{b} & \colhead{${\rm V_{LSR}}$} & \colhead{Flux} 
& \colhead{${\rm DGR_{Cloud}\tablenotemark{a}}$}
& \colhead{${\rm DGR_{Cloud}\tablenotemark{a}}$} 
& \colhead{$\rho$} \\
\colhead{} & \colhead{[deg]} & \colhead{[deg]} & \colhead{${\rm (km ~s^{-1})}$} & \colhead{(Jy km s$^{-1}$)} & {${100\mu {\rm m}/ 21 {\rm cm}}$} & {${60\mu {\rm m} / 21 {\rm cm}}$} & \colhead{}  
}
\startdata
{\bf HVCs} &   &   &  & &   & & \\ 
AC & 189 & $-20$ & $-116$ & $5.6 \times 10^3 $& $11 \pm 17$  & $2.1 \pm 3.8$ & 0.94\\ 
G  & 84  & $-13$ & $-94$  & $5.1 \times 10^3 $ & $-6.3 \pm 6.6$ & $0.5 \pm 2.8$ & 0.72 \\
WA & 212 &  24   &   89   & $3.0 \times 10^2 $& $4.7 \pm 6.7$  &  $-0.75 \pm 2.5$ & 0.72\\
P  & 128 & $-34$ & $-392$ & $3.4 \times 10^2 $& $-8.3 \pm 8.7$ & $-0.1 \pm 2.9$ & 0.74 \\
M  & 198 &  50 & $-85$    & $1.6 \times 10^3 $ &  $13 \pm 3.4$  & $5.1 \pm 1.6$ & 0.42 \\
C  & 34 & 18 & $-125$   & $8.2 \times 10^2 $ &  $-26 \pm 16$  & $-3.9 \pm 5.3$ & 0.90 \\
\hline
{\bf LVCs} &   &   &  & &   & & \\ 
L1 &   84  & $-13$  &  20  &  $5.9 \times 10^2$    &  $39 \pm 20$  & $12 \pm 5.6$ & 0.81 \\
L2 & 252 &  60   &  48    & $6.6 \times 10^3$  & $17 \pm 4.4$  & $6.4 \pm 2.0$ & 0.81 \\
L3 & 322 &  68   &  30  & $ 3.0 \times 10^2$  & $172 \pm 16$  & $64 \pm 7.8 $ & 0.86 \\
L4 & 5   &  53  & $-60$ & $2.3 \times 10^2$  & $153 \pm 19$  & $45 \pm 7.3$ & 0.86\\
L5 & 253   &  60  & $53$ & $3.1 \times 10^2$  & $7.2 \pm 9.1$  & $5.9 \pm 4.6$ & 0.40 \\
L6 & 32 &  84  & $-58$ & $1.7 \times 10^2$  & $2.8 \pm 6.3$  & $1.9 \pm 3.2$ & 0.59\\
L7 & 47 &  39  & $-60$ & $2.7 \times 10^3$  & $67 \pm 16$  & $27 \pm 4.0$ & 0.91 \\
L8 & 147 &  $-29$  & $43$ & $2.9 \times 10^2$  & $27 \pm 18$  & $21 \pm 7.1$ & 0.72 \\
L9 &  38 & 65 & $-85$ & $2.4 \times 10^3$  & $34 \pm 4.0$  & $11 \pm 1.7$ & 0.91 \\
\hline
{\bf Dwarfs} &   &   &  & &   & & \\ 
Leo T &  214 & 43 & $30$ & $7 \times 10^0$  & $8.5 \pm 13$  & $-6.8 \pm 5.4 $ & 0.54 \\
\hline
\enddata
\tablenotetext{a}{All dust-to-gas ratio values are listed in ${\rm 10^{-22} MJy sr^{-1} ~cm^2}$.}
\tablecomments{A table of objects observed in our preliminary survey of HVCs, LVCs, and a dwarf Galaxy. $\rho$ is the measured correlation coefficient between the errors in the 60 and 100 $\mu$m DGRs. Errors are 1-$\sigma$ values. \label{DGR_tab}}
\end{deluxetable}
\end{center}

\begin{figure*}
\begin{center}
\includegraphics[scale=.60]{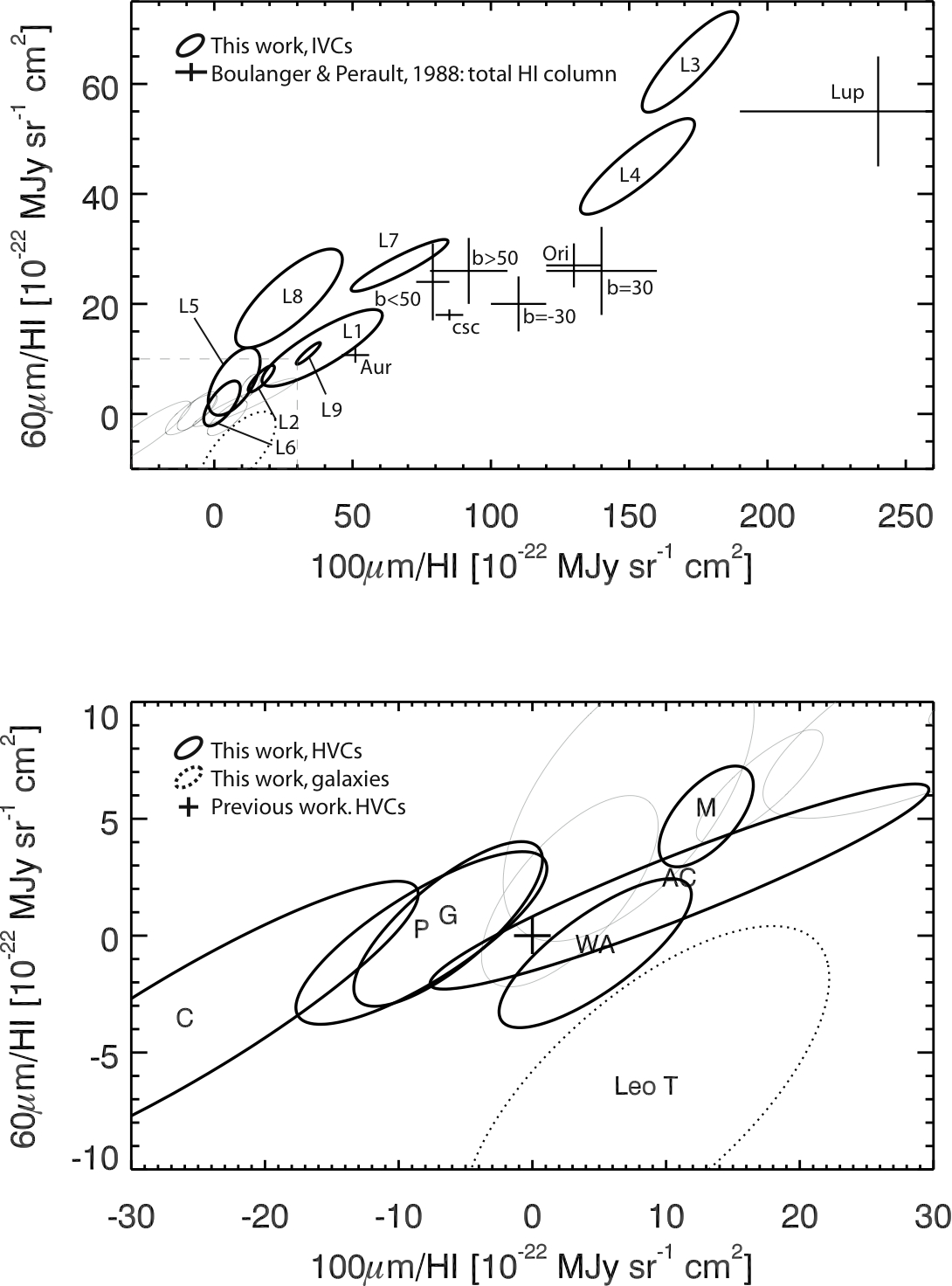}
\caption{Two plots of the results of our analysis for 6 HVCs and 9 LVCs and the Leo T dwarf. The top plot focuses on the LVCs, the bottom plot focuses on the HVCs and the Leo T dwarf. HVCs are labeled by the complex to which they belong. Our data are shown by 1-$\sigma$ ellipses, which take into account the correlation, $\rho$, in errors between the 60$\mu$m and 100$\mu$m DGR measurements. In the top plot we have overplotted data from BP88  - this is the DGR for the total \hi column from 3 regions of the sky (Auriga, Orion and Lupus) as well as the Northern and Southern Galactic Caps beyond $|b| = 50$, the sky near $|b| = 30$ and the fit to an assumed $csc$(b) distribution. We also show the expected zero infrared emission for HVCs with a cross in the bottom plot. Due to lack of available DIRBE calibration the BP88 data may have systematic errors at the 20\% level. Note the detection of the dust in complex M, the non-detection of dust in L5 and L6, as well as the increased DGR$_{60\mu \rm{m}}$/DGR$_{100\mu \rm{m}}$ in LVCs compared to the total column results. \label{DGR_plot}}
\end{center}
\end{figure*}

\section{Results}\label{s_results}

\subsection{HVCs}\label{hvcres}
The results for HVCs, as shown in Figure \ref{DGR_plot} and Table \ref{DGR_tab}, primarily confirm the result from WB86 --- HVCs have no detectable dust emission. Some of these HVCs (WA, G, and P) have relatively small errors, clearly setting them apart from the typical measured DGR in the Galactic disk, $D_{\lambda, 60} \sim 20 \times {\rm 10^{-22} MJy~sr^{-1} ~cm^2}$ and $D_{\lambda, 100} \sim 100 \times {\rm 10^{-22} MJy~sr^{-1} ~cm^2}$ (\citei{BP88}). The results are less clear for the HVCs we sampled in complexes C and AC, which have larger error bars. Note that our measurements do not strictly rule out the values found in \citet{M-D05}. Also note that a few of the HVCs show negative DGR at low significance ($\le 2 \sigma$),which we take to be measurement noise. The most surprising result is a $> 3\sigma$ \emph{detection} of dust in a cloud in complex M. It seems that complex M does indeed have some small amount of heated dust. Complex M is a relatively low-velocity HVC complex and is on the border of the HVC/IVC classification. Indeed, some think that the cloud may be related to the IVC arch at a distance of $\sim 1$ kpc (\citei{vWW04}, \citei{Wakker2001}). Complex M has been shown to be within 4 kpc by UV absorption line measurements \citep{Danly93} and Ca II K absorption (e.g. \citei{Keenan95}), but does not have significant lower limits.  If it were indeed related to the IVC arch, that might explain the detection of dust, given the strong expected UV flux near the plane of the Galaxy (e.g. \citei{Putman03}, \citei{B-HM99}) and known solar metallicity of the IVC Arch \citep{Richter01}. There are no absorption-line metallicity measurements in the literature as there has not been a high-resolution measurement of \hi towards the stars detected behind complex M. Measurements of [S {\sc II}] and H$\alpha$ by \citet{Tufte98} point to a metallicity near solar, but there are significant errors in this measurement. A solar metallicity would be consistent with a Galactic origin for complex M. All other measured HVCs (WA, G, P and AC) are consistent with low dust dust content and/or negligible ISRF at the cloud.

\subsection{LVCs}\label{lvcres}
Results for the nine LVCs measured are also shown on the top of Figure \ref{DGR_plot} (ellipses) and Table \ref{DGR_tab}. Our measurements of LVCs span a very large range in both DGR$_{100}$ and DGR$_{60}$. We find 2 such clouds, L5 and L6, which are consistent with zero DGR and inconsistent with normal disk gas. These are candidate LVHCs. We also find one cloud (L2) with DGR consistent with the DGR found in the complex M detection. L2 may be consistent with whatever kind of object complex M is (a nearby HVC or a fast-moving disk cloud), or it may be on the far end of the typical range of LVC DGRs. The remainder of the LVCs are distributed over a wide range of DGR values. These values roughly span the same range shown in the analysis by \citet{BP88} of the DGR of various areas of the high galactic latitude sky, including star-forming regions, the Galactic caps and the overall sky averages beyond b$=|30|$: $10 \le $DGR$_{60} \le 55$, $50 \le $DGR$_{100} \le 240$, in units of ${\rm 10^{-22} MJy/sr ~cm^2}$.

Of note is that the measured DGRs in these LVCs have a typically higher DGR$_{60}$/DGR$_{100}$ value than the measurements from \citet{BP88}. There are a number of possible reasons for this effect. The effect may arise from an overall smaller grain size distribution, which would result in a higher temperature for the grains and thus `bluer' grains. There also may be a larger number of very-small grains (VSGs); VSGs are stochastically heated by UV photons and thereby contribute significantly to the 60$\mu m$ grain emission. Given the conclusion in \citet{Fitzpatrick96} that LVCs have less material locked up in grains due to grain destruction from shocks, the decrease in average grain size (or increase in VSGs) may indeed be the correct interpretation of these results. This result is consistent with previous work on the subject by \citet{HRK88}, but inconsistent with observations by \citet{DB90}, who find a lower DGR$_{60}$/DGR$_{100}$ value in such clouds. 

We find that it is very difficult to make statistically sound claims about the total number of LVHCs that exist based on these data, primarily due to the complex selection effects involved. We might argue that if we take the number of LVHCs observed and divide out both our detection efficiency for LVHCs and the area of the sky surveyed we might be able to extrapolate to the total number of LVHCs in the sky. There are a few difficulties in pursuing this line of reasoning. The first is that we detect a very small number of LVHCs. With only two clear cases of candidate LVHCs, we have very large error bars simply from Poisson statistics. It is also difficult to determine how many `clouds' each LVHC comprises; to compare to the the WVW91 catalog we need to have the same criteria for determining cloud identity. We also encounter the problem of not having a very accurate measurement of our detection efficiency - certainly LVHCs at $V_{\rm LSR}$ = 0 are undetectable to this survey, but at what velocity they become detectable varies significantly over the area of the sky observed. The detection efficiency is also strongly dependent upon the quality and depth of the GALFA-\hi data and IRAS data, which vary significantly. It is also important to note that since the area of the sky surveyed is rather small and that HVCs are known to be significantly clustered on the sky, any extrapolation may be strongly influenced by whether the area of the sky covered happens to intersect an LVHC complex. We therefore leave a statistical analysis of the LVHC population to a future paper with a larger cloud sample and quantifiable selection criteria. 

\subsection{Leo T}\label{ltres}
We find that Leo T has no detectable dust emission in either 60$\mu$m or 100$\mu$m, consistent with our conjecture that a gas-rich dwarf galaxy near the Milky Way will have cooler, or less, dust than the Milky Way itself. This finding is also consistent with the analysis in \citet{LF98}, who show that gas-rich dwarf galaxies have significantly lower dust fractions (by mass) that the Milky Way. This result leads us to believe that the DGR method may also be useful for determining whether a compact HI cloud is simply a cloud in the Galactic disk, or whether it is a gas-rich dwarf Galaxy candidate. 

\section{Conclusion}\label{s_conc}
We have shown that the GALFA-\hi data and IRIS reduction of the IRAS data make an excellent match for use in determining the DGR of neutral clouds in our Galaxy (\S \ref{s_obs}). We showed that the simple least-squares (or chi-squared) fit method for determining the dust emission from neutral clouds with a distinct velocity (LVCs \& HVCs) is flawed in circumstances where FIR fluxes are low compared to that of the zero-velocity gas (\S \ref{stanap}). We developed the displacement-map method for determining the DGR of such neutral clouds and showed it to be an accurate and robust technique (\S \ref{dmap}; \S \ref{tests}). We applied this method to the 9 LVCs and 6 HVCs found in the GALFA-\hi survey area mapped to date along with the Leo T dwarf galaxy (\S \ref{targ_sel}). Only one of the 6 HVCs, a cloud from complex M, showed a significant detection of dust. This cast into doubt the distance to, and history of, complex M, which may be related to more nearby, disk-associated clouds such as the Intermediate-Velocity Arch (\S \ref{hvcres}). Two lower-velocity clouds, L5 and L6, were shown to have no detectable dust content, and we therefore consider them candidate LVHCs (\S \ref{lvcres}). We also find that most LVCs and IVCs have significantly hotter dust than typical Milky Way, as measured by their DGR$_{60}$/DGR$_{100}$ value, which we find to be consistent with observations of decreased dust depletion in LVCs. We also find that Leo T has no detectable dust emission, as we expect for a dwarf galaxy with relatively low ISRF (\S \ref{ltres}).

The clearest follow-up work needed here is to examine larger swathes of the Arecibo-accessible sky for HVCs and LVCs. As the GALFA-\hi survey accumulates data in the coming years we expect to measure many more clouds using this method to determine if there are more LVHC candidates, if any other HVCs show clear signs of dust, and if the trend of hotter LVC dust remains. If a large enough sample of LVC measurements can be built, we may be able to statistically address the question of how many LVHCs there are as compared to HVCs, thus giving us a new discriminant between HVC models (see Figure \ref{model_plot}). In particular, we are pursuing automated methods of searching for LVCs in the GALFA-\hi survey, which will allow us to determine our selection biases much more accurately and thus extrapolate to the full LVHC population. These cloud isolation methods are already showing significant promise \citep{HP08}.

Expanding upon our one-object proof-of-concept analysis of Leo T, we also intend to do a similar search for dwarf-like clouds in the GALFA-\hi survey data. Any compact clouds lacking associated infrared flux would be very interesting candidates for optical follow up as gas-rich local-group dwarf candidates.

It is also crucial to determine whether the conjecture that motivated this work, that lack of infrared emission in clouds is correlated to halo membership, is indeed correct. Now that we have at least two LVCs with no discernible dust, we wish to determine whether they indeed lie significantly beyond the disk of our Galaxy ($|z| > 1$ kpc). Observations are underway to examine stars in the lower halo along the line of sight to these objects to look for absorption features. In addition, we would like to examine other LVCs that do exhibit infrared emission, to determine whether they are near the disk of the Galaxy, as we would expect from their dust emission.

JEGP would like to thank Christopher McKee, Evan Levine, Louis-Benoit Desroches and Robert Goldston for many helpful conversations. JEGP would also like to thank Suzanne McKee for insights into parallels to the computer vision community. JEGP was supported in part by The National Science Foundation grants AST-0709347 and AST-0406987. CH was supported in part by The National Science Foundation grant AST-0406987. MEP was supported by NSF grant AST-0707597. The authors would like to thank the ALFALFA team and the staff of the Arecibo Observatory, without whose hard work this would not have been possible. The Arecibo Observatory is part of the National Astronomy and Ionosphere Center, which is operated by Cornell University under a cooperative agreement with the National Science Foundation. 

\bibliographystyle{apj}

\begin{thebibliography}{39}
\expandafter\ifx\csname natexlab\endcsname\relax\def\natexlab#1{#1}\fi

\bibitem[{{Beichman}(1987)}]{Beichman87}
{Beichman}, C.~A. 1987, \araa, 25, 521

\bibitem[{{Bland-Hawthorn} \& {Maloney}(1999)}]{B-HM99}
{Bland-Hawthorn}, J., \& {Maloney}, P.~R. 1999, in Astronomical Society of the
  Pacific Conference Series, Vol. 166, Stromlo Workshop on High-Velocity
  Clouds, ed. B.~K. {Gibson} \& M.~E. {Putman}, 212

\bibitem[{{Boulanger} \& {Perault}(1988)}]{BP88}
{Boulanger}, F., \& {Perault}, M. 1988, \apj, 330, 964

\bibitem[{{Bregman}(1980)}]{Bregman1980}
{Bregman}, J.~N. 1980, \apj, 236, 577

\bibitem[{{Danly} {et~al.}(1993){Danly}, {Albert}, \& {Kuntz}}]{Danly93}
{Danly}, L., {Albert}, C.~E., \& {Kuntz}, K.~D. 1993, \apjl, 416, L29

\bibitem[{{Deul} \& {Burton}(1990)}]{DB90}
{Deul}, E.~R., \& {Burton}, W.~B. 1990, \aap, 230, 153

\bibitem[{{Fitzpatrick}(1996)}]{Fitzpatrick96}
{Fitzpatrick}, E.~L. 1996, \apjl, 473, L55

\bibitem[{{Gillmon} \& {Shull}(2006)}]{GS06}
{Gillmon}, K., \& {Shull}, J.~M. 2006, \apj, 636, 908

\bibitem[{{Gillmon} {et~al.}(2006){Gillmon}, {Shull}, {Tumlinson}, \&
  {Danforth}}]{Gillmon06}
{Gillmon}, K., {Shull}, J.~M., {Tumlinson}, J., \& {Danforth}, C. 2006, \apj,
  636, 891

\bibitem[{{Giovanelli} {et~al.}(2005){Giovanelli}, {Haynes}, {Kent},
  {Perillat}, {Saintonge}, {Brosch}, {Catinella}, {Hoffman}, {Stierwalt},
  {Spekkens}, {Lerner}, {Masters}, {Momjian}, {Rosenberg}, {Springob},
  {Boselli}, {Charmandaris}, {Darling}, {Davies}, {Lambas}, {Gavazzi},
  {Giovanardi}, {Hardy}, {Hunt}, {Iovino}, {Karachentsev}, {Karachentseva},
  {Koopmann}, {Marinoni}, {Minchin}, {Muller}, {Putman}, {Pantoja}, {Salzer},
  {Scodeggio}, {Skillman}, {Solanes}, {Valotto}, {van Driel}, \& {van
  Zee}}]{Giovanelli05}
{Giovanelli}, R., {Haynes}, M.~P., {Kent}, B.~R., {Perillat}, P., {Saintonge},
  A., {Brosch}, N., {Catinella}, B., {Hoffman}, G.~L., {Stierwalt}, S.,
  {Spekkens}, K., {Lerner}, M.~S., {Masters}, K.~L., {Momjian}, E.,
  {Rosenberg}, J.~L., {Springob}, C.~M., {Boselli}, A., {Charmandaris}, V.,
  {Darling}, J.~K., {Davies}, J., {Lambas}, D.~G., {Gavazzi}, G., {Giovanardi},
  C., {Hardy}, E., {Hunt}, L.~K., {Iovino}, A., {Karachentsev}, I.~D.,
  {Karachentseva}, V.~E., {Koopmann}, R.~A., {Marinoni}, C., {Minchin}, R.,
  {Muller}, E., {Putman}, M., {Pantoja}, C., {Salzer}, J.~J., {Scodeggio}, M.,
  {Skillman}, E., {Solanes}, J.~M., {Valotto}, C., {van Driel}, W., \& {van
  Zee}, L. 2005, \aj, 130, 2598

\bibitem[{{Grevich} \& {Putman}(2008)}]{GP08}
{Grevich}, J., \& {Putman}, M. 2008, in prep

\bibitem[{{Heiles} {et~al.}(1988){Heiles}, {Reach}, \& {Koo}}]{HRK88}
{Heiles}, C., {Reach}, W.~T., \& {Koo}, B.-C. 1988, \apj, 332, 313

\bibitem[{{Hsu} \& {Putman}(2008)}]{HP08}
{Hsu}, W.-H., \& {Putman}, M.~E. 2008, in prep

\bibitem[{{Jefferys}(1980)}]{Jefferys80}
{Jefferys}, W.~H. 1980, Astron J., 85

\bibitem[{{Kalberla} {et~al.}(2005){Kalberla}, {Burton}, {Hartmann}, {Arnal},
  {Bajaja}, {Morras}, \& {P{\"o}ppel}}]{Kalberla2005}
{Kalberla}, P.~M.~W., {Burton}, W.~B., {Hartmann}, D., {Arnal}, E.~M.,
  {Bajaja}, E., {Morras}, R., \& {P{\"o}ppel}, W.~G.~L. 2005, \aap, 440, 775

\bibitem[{{Keenan} {et~al.}(1995){Keenan}, {Shaw}, {Bates}, {Dufton}, \&
  {Kemp}}]{Keenan95}
{Keenan}, F.~P., {Shaw}, C.~R., {Bates}, B., {Dufton}, P.~L., \& {Kemp}, S.~N.
  1995, \mnras, 272, 599

\bibitem[{{Lisenfeld} \& {Ferrara}(1998)}]{LF98}
{Lisenfeld}, U., \& {Ferrara}, A. 1998, \apj, 496, 145

\bibitem[{{Maller} \& {Bullock}(2004)}]{MB2004}
{Maller}, A.~H., \& {Bullock}, J.~S. 2004, \mnras, 355, 694

\bibitem[{{Mayer} {et~al.}(2006){Mayer}, {Mastropietro}, {Wadsley}, {Stadel},
  \& {Moore}}]{Mayer06}
{Mayer}, L., {Mastropietro}, C., {Wadsley}, J., {Stadel}, J., \& {Moore}, B.
  2006, \mnras, 369, 1021

\bibitem[{{Miville-Desch{\^e}nes} {et~al.}(2005){Miville-Desch{\^e}nes},
  {Boulanger}, {Reach}, \& {Noriega-Crespo}}]{M-D05}
{Miville-Desch{\^e}nes}, M.-A., {Boulanger}, F., {Reach}, W.~T., \&
  {Noriega-Crespo}, A. 2005, \apjl, 631, L57

\bibitem[{{Miville-Desch{\^e}nes} \& {Lagache}(2006)}]{M-DL06}
{Miville-Desch{\^e}nes}, M.-A., \& {Lagache}, G. 2006, in Astronomical Society
  of the Pacific Conference Series, Vol. 357, Astronomical Society of the
  Pacific Conference Series, ed. L.~{Armus} \& W.~T. {Reach}, 167

\bibitem[{{Muller} {et~al.}(1963){Muller}, {Oort}, \& {Raimond}}]{Muller1963}
{Muller}, C.~A., {Oort}, J.~H., \& {Raimond}, E. 1963, C. R. Acad. Sci. Paris,
  257, 1661

\bibitem[{{Odegard} {et~al.}(2007){Odegard}, {Arendt}, {Dwek}, {Haffner},
  {Hauser}, \& {Reynolds}}]{Odegard07}
{Odegard}, N., {Arendt}, R.~G., {Dwek}, E., {Haffner}, L.~M., {Hauser}, M.~G.,
  \& {Reynolds}, R.~J. 2007, \apj, 667, 11

\bibitem[{{Oort}(1966)}]{oort66}
{Oort}, J.~H. 1966, \bain, 18, 421

\bibitem[{{Peek} \& {Heiles}(2008)}]{PH08}
{Peek}, J.~E.~G., \& {Heiles}, C. 2008, ArXiv e-prints

\bibitem[{{Peek} {et~al.}(2008){Peek}, {Putman}, \& {Sommer-Larsen}}]{PPS-L08}
{Peek}, J.~E.~G., {Putman}, M.~E., \& {Sommer-Larsen}, J. 2008, \apj, 674, 227

\bibitem[{{Putman} {et~al.}(2003){Putman}, {Bland-Hawthorn}, {Veilleux},
  {Gibson}, {Freeman}, \& {Maloney}}]{Putman03}
{Putman}, M.~E., {Bland-Hawthorn}, J., {Veilleux}, S., {Gibson}, B.~K.,
  {Freeman}, K.~C., \& {Maloney}, P.~R. 2003, \apj, 597, 948

\bibitem[{{Richter} {et~al.}(2001){Richter}, {Sembach}, {Wakker}, {Savage},
  {Tripp}, {Murphy}, {Kalberla}, \& {Jenkins}}]{Richter01}
{Richter}, P., {Sembach}, K.~R., {Wakker}, B.~P., {Savage}, B.~D., {Tripp},
  T.~M., {Murphy}, E.~M., {Kalberla}, P.~M.~W., \& {Jenkins}, E.~B. 2001, \apj,
  559, 318

\bibitem[{Rosenfeld(1969)}]{Rosenfeld69}
Rosenfeld, A. 1969, ACM Comput. Surv., 1, 147

\bibitem[{{Schlegel} {et~al.}(1998){Schlegel}, {Finkbeiner}, \&
  {Davis}}]{SFD98}
{Schlegel}, D.~J., {Finkbeiner}, D.~P., \& {Davis}, M. 1998, \apj, 500, 525

\bibitem[{{Sommer-Larsen}(2006)}]{SL06}
{Sommer-Larsen}, J. 2006, \apjl, 644, L1

\bibitem[{{Stanimirovi{\'c}} {et~al.}(2006){Stanimirovi{\'c}}, {Putman},
  {Heiles}, {Peek}, {Goldsmith}, {Koo}, {Kr{\v c}o}, {Lee}, {Mock}, {Muller},
  {Pandian}, {Parsons}, {Tang}, \& {Werthimer}}]{Stanimirovic2006}
{Stanimirovi{\'c}}, S., {Putman}, M., {Heiles}, C., {Peek}, J.~E.~G.,
  {Goldsmith}, P.~F., {Koo}, B.-C., {Kr{\v c}o}, M., {Lee}, J.-J., {Mock}, J.,
  {Muller}, E., {Pandian}, J.~D., {Parsons}, A., {Tang}, Y., \& {Werthimer}, D.
  2006, \apj, 653, 1210

\bibitem[{{Tufte} {et~al.}(1998){Tufte}, {Reynolds}, \& {Haffner}}]{Tufte98}
{Tufte}, S.~L., {Reynolds}, R.~J., \& {Haffner}, L.~M. 1998, \apj, 504, 773

\bibitem[{{van Woerden} \& {Wakker}(2004)}]{vWW04}
{van Woerden}, H., \& {Wakker}, B.~P. 2004, in ASSL Vol. 312: High Velocity
  Clouds, ed. H.~{van Woerden}, B.~P. {Wakker}, U.~J. {Schwarz}, \& K.~S. {de
  Boer}, 195

\bibitem[{{Wakker}(1991)}]{wakker1991}
{Wakker}, B.~P. 1991, in IAU Symposium, Vol. 144, The Interstellar Disk-Halo
  Connection in Galaxies, ed. H.~{Bloemen}, 27--40

\bibitem[{{Wakker}(2001)}]{Wakker2001}
{Wakker}, B.~P. 2001, \apjs, 136, 463

\bibitem[{{Wakker}(2004)}]{Wakker2004}
{Wakker}, B.~P. 2004, in ASSL Vol. 312: High Velocity Clouds, ed. H.~{van
  Woerden}, B.~P. {Wakker}, U.~J. {Schwarz}, \& K.~S. {de Boer}, 25

\bibitem[{{Wakker} \& {Boulanger}(1986)}]{WB1986}
{Wakker}, B.~P., \& {Boulanger}, F. 1986, \aap, 170, 84

\bibitem[{{Wakker} \& {van Woerden}(1991)}]{WVW1991}
{Wakker}, B.~P., \& {van Woerden}, H. 1991, \aap, 250, 509

\end{thebibliography}

\end{document}